\documentclass[options]{article}

\usepackage{geometry}
\newcommand{\be}{\begin{equation}}

\newcommand{\ee}{\end{equation}}

\newcommand{\ba}{\begin{eqnarray}}
\newcommand{\ea}{\end{eqnarray}}
\newcommand{\baa}{\begin{eqnarray*}}
\newcommand{\eaa}{\end{eqnarray*}}
\newcommand{\bb}{\begin{thebibliography}}
\newcommand{\eb}{\end{thebibliography}}

\newcommand{\lab}[1]{\label{#1}}
\newcommand{\re}[1]{(\ref{#1})}

\newtheorem{pr}{Proposition}

\newtheorem{lem}{Lemma}

\begin{document}

%\begin{titlepage}
\vspace*{2mm}

\begin{center}

{\Large \bf Toda-Schr\"odinger correspondence and orthogonal polynomials}

\vspace{15mm}

{\large \bf Satoshi TSUJIMOTO}

\medskip

{\it Department of Applied Mathematics and Physics, Graduate School of Informatics,
Kyoto University, Kyoto 606-8501, Japan}

\medskip

and

\medskip

{\large \bf Alexei ZHEDANOV}

\medskip

{\em Donetsk Institute for Physics and Technology, Donetsk 83114,  Ukraine}

\end{center}

%\vspace*{5mm}

\begin{abstract}
It is known that the unrestricted Toda chain is equivalent
to the Riccati equation for the Stieltjes function of the orthogonal polynomials. 
Under a special condition, this 
Riccati equation can be reduced to the Schr\"odinger equation. We show that this condition is equivalent to type B solutions of the Toda chain. We establish some nontrivial consequences arising from this Toda-Schr\"odinger correspondence. In particular, we show that the KdV densities can be identified with the moments of the corresponding orthogonal polynomials. We establish equivalence between type B solutions of the Toda molecule and the Bargmann potentials of the Schr\"odinger equation

\vspace{2cm}


\vspace{2cm}


\end{abstract}

\bigskip\bigskip

\section{Introduction}
The purpose of this paper is establishing of nontrivial relations between solutions of the Schr\"odinger equation and so-called type B solutions of the unrestricted Toda chain.  The main tool of our approach is a  connection between solutions of the unrestricted Toda chain and orthogonal polynomials proposed in \cite{PSZ}. Although today solutions of the Toda chain are well studied, still there are interesting relations with quantum mechanics which we are going to present here.

The paper is organized as follows.

In the second section we recall relations between the Toda chain and orthogonal polynomials based mostly on the result of the paper \cite{PSZ}. We find the condition under which equation for the Stieltjes function can be presented in the form of the Standard Schr\"odinger equation.  

In the third section we show that this condition is equivalent to a specific "mirror" boundary conditions for the Toda chain. In turn, this is equivalent to solutions of type B of the Toda chain introduced by Ueno and Takasaki \cite{UeTa}.

In the fourth section, we show that there is one-to-one correspondence between conserved densities $\sigma_n(x)$ of the Korteweg-de Vries equation and the moments  $c_n(t)$ of orthogonal polynomials corresponding to the type B Toda chain solutions.  

In the fifth section, the finite-dimensional case (i.e. Toda chain molecule) is considered. The main result of this section is establishing the equivalence between the class of the Bargmann (reflectionless) potentials of the Schr\"odinger equation and the type B Toda molecule. 

In the sixth section, we consider spectral problems (direct and inverse) for finite Jacobi matrices corresponding to the type B and C. Matrices with such (and similar) structures are important in applications, e.g. in perfect state transfer in quantum informatics \cite{VZ_PST}.  

In the seventh section, special elementary solutions of Toda chain of type B are considered. These solutions correspond to the well known exactly solvable potential of the Schr\"odinger equation. In turn, solutions of the Toda chain correspond to some classical orthogonal polynomials. 

Finally, in the eight section, we consider an example of the Schr\"odinger equation with the linear potential. This leads to rational solutions of the Painlev\'e-II equation.

\section{Toda chain, Stieltjes function and orthogonal polynomials}
Let $c_0(t)$ and $u_0(t)$ be two arbitrary analytic functions in $t$.
Define $b_0(t) = \dot c_0/c_0$ and then construct functions
$u_n(t), \; n= \pm 1, \pm 2, \dots$ and $b_n(t), \; n= \pm 1, \pm
2, \dots$ by using the Toda chain equations \cite{Toda} \be \dot
b_k = u_{k+1}-u_{k}, \quad \dot u_k = u_k (b_{k} - b_{k-1})
\lab{Toda_eq}. \ee It is clear that all these functions $u_n(t),
b_n(t)$ are determined uniquely if one assumes that $u_n(t) \ne
0$.

We can construct another sequence of functions $c_n(t), \;
n=1,2,\dots$ uniquely from the nonlinear recurrence
relation \be c_1 = \dot c_0, \quad   c_{n+1} = \dot c_n +
\frac{u_0}{c_0} \sum_{s=0}^{n-1} c_s c_{n-1-s}, \; n=1,2,\dots.
\lab{chain_mom} \ee

Let us construct also the third sequence of functions $H_n(t), \: n=0, \pm 1, \pm 2,\dots$ uniquely from the nonlinear recurrence
relation
\be
\frac{d^2}{dt^2} \log(H_n) + u_0 = \frac{H_{n-1}H_{n+1}}{H_n^2},
\; n=1,2,\dots \lab{Toda_H} \ee
with initial conditions $H_0=1, \: H_1(t) = c_0(t)$.

It appears that all these relations are equivalent in case if $u_n(t) \ne 0, \: n=0,1,2,\dots$ \cite{PSZ}.
Moreover, the functions $H_n(t)$ can be related with moments $c_n(t)$ as
\be
H_n(t) = det||c_{i+k}(t)||_{i,k=0}^{n-1},
\lab{H_n_c} \ee
{i.e.\ }$H_n(t)$ are the Hankel determinants corresponding to the moments $c_n(t)$.
Relations between $H_n(t)$ and the recurrence coefficients $u_n(t), b_n(t)$ are
\be u_n =
\frac{H_{n-1}H_{n+1}}{H_n^2}, \quad b_n = \frac{d }{d t} \log(H_{n+1}/H_n) \lab{ub_H}. \ee
Thus nondegenerate condition $u_n(t) \ne 0$ is equivalent to the condition $H_n(t) \ne 0$

Let $F(z;t)$ be the Stieltjes functions, i.e.
a formal generating function corresponding to these moments: \be
F(z;t) = \sum_{n=0}^{\infty} c_n(t) z^{-n-1} \lab{St_F}. \ee It is
easily verified \cite{PSZ} that relations \re{chain_mom} are equivalent to
the Riccati equation for the Stieltjes function \be \dot F = - c_0
+z F - u_0 F^2/c_0 \lab{Ric_F1}. \ee

Using the substitution \be F(z;t) = \frac{c_0(t) \dot
\psi(z;t)}{u_0(t) \psi(z;t)} + B(z;t), \lab{F_psi} \ee where
$$
2B(z;t) = z \frac{c_0(t)}{u_0(t)} + \frac{d}{dt} \left(
\frac{c_0(t)}{u_0(t)}\right),
$$
we transform the Riccati equation to the Sturm-Liouville equation
\be \ddot \psi + (u_0 -\dot b_{-1}/2 -(b_{-1}-z)^2/4) \psi =0,
\lab{schr_toda} \ee
 where $b_{-1}(t)=b_0 - \dot u_0 /u_0$ as assumed by the Toda chain
 equations \re{Toda_eq}.

Conversely, starting from an appropriate solution of the
Sturm-Liouville equation \re{schr_toda} we can construct the
Stieltjes function $F(z;t)$ satisfying the Riccati equation
\re{Ric_F1} and then reconstruct corresponding moments $c_n(t), \:
n=1,2,3,\dots$.

With the moments $c_n(t)$ one can associate the monic orthogonal
polynomials $P_n(x;t)$ by the formulas \ba
P_n(x;t)=\frac{1}{H_n(t)}\left |
\begin{array}{cccc} c_0(t) & c_{1}(t) & \dots &
c_{n}(t)\\ c_{1}(t)& c_{2}(t) & \dots & c_{n+1}(t)\\ \dots & \dots & \dots & \dots\\
c_{n-1}(t) & c_{n}(t) & \dots & c_{2n-1}(t) \\ 1 & x & \dots & x^n
\end{array} \right |. \lab{P_det} \ea
Then
$P_n(x;t)$ are polynomials of exact degree $n$ satisfying
three-term recurrence relation \be P_{n+1} + b_n P_n + u_n P_{n-1}
= x P_n \lab{rec_P_n} \ee with the initial conditions \be P_0(x;t) =1,
\; P_1(x;t) = x- b_0(t) \lab{ini_P_n}. \ee

Moreover, the orthogonal polynomials $P_n(x;t)$ satisfy the
relation \be \dot P_n(x;t) = -u_n P_{n-1}(x;t) + u_0
P^{(1)}_{n-1}(x;t), \lab{dot_P_n} \ee where $P^{(1)}_{n}(x;t)$ are
so-called associative polynomials defined by the recurrence
relation \be P^{(1)}_{n+1}(x;t) + b_{n+1}(t) P^{(1)}_{n}(x;t) +
u_{n+1}(t) P^{(1)}_{n-1}(x;t) = x P^{(1)}_{n}(x;t), \lab{rec_ass}
\ee with initial conditions
$$
P^{(1)}_{0}(x;t) =1, \; P^{(1)}_{1}(x;t)= x- b_1(t).
$$
We have \be F(x) P_n(x) - P_{n-1}^{(1)}(x) = F_n(x), \lab{F_n_def}
\ee where $F_n(x) = h_n x^{-n-1} + O(x^{-n-2})$ are functions of
the second kind satisfying the same recurrence relation \be
F_{n+1}(x) + b_n F_n(x) + u_n F_{n-1}(x) = x F_n(x), \;
n=1,2,\dots  \lab{rec_F_n}, \ee that the polynomials $P_n(x)$.
Clearly $F_0(x)=F(x)$. For $n=0$ relation \re{rec_F_n} looks as
\cite{PSZ}:
$$
F_1(x) + b_0 F(x) +1 = xF(x).
$$
The moments $c_n(t)$ define a linear functional $\sigma(t)$ acting
on the space of polynomials by its values on the monomials. \be
\langle \sigma(t), x^n \rangle = c_n(t) \lab{def_sigma} \ee
Polynomials $P_n(x;t)$ are orthogonal with respect to the
functional $\sigma$: \be \langle \sigma(t), P_n(x;t) P_m(x;t)
\rangle = h_n(t) \: \delta_{nm}, \lab{ort_P} \ee where
$$
h_n(t) = \frac{H_{n+1}(t)}{H_n(t)} = c_0 u_1(t) u_2(t) \dots u_n(t).
$$

Thus the functions $u_0(t), c_0(t)$ (or, equivalently, $u_0(t),
b_0(t) = \dot c_0/c_0$) generate uniquely a set of orthogonal
polynomials $P_n(x;t)$ and a linear functional $\sigma(t)$
providing orthogonality of these polynomials.

Assume that the functions $u_0(t), c_0(t)$ satisfy the condition
\be u_0(t) = \kappa \: c_0(t) \lab{u0c0}, \ee with a constant
$\kappa$ not depending on $t$. 

Condition \re{u0c0} is equivalent to the condition
$b_{-1}(t) = 0$. Indeed,
$$
b_{-1} = \frac{d \log (c_0/u_0)}{dt} =0.
$$
Under this condition, the Sturm-Liouville equation \re{schr_toda} is reduced to the
standard Schr\"odinger equation \be \ddot \psi + (u_0(t) -z^2/4)
\psi =0 \lab{shcr}, \ee where $u_0(t)$ plays the role of the
"potential" of the Schr\"odinger equation and $z^2/4$ is the
"energy". Note that the constant $\kappa$ can be chosen equal to
1. Indeed, the Toda chain equations for the moments \re{chain_mom}
are preserved under the scaling transform $c_n \to \mu c_n, \:
n=0,1,2,\dots$ and $u_0 \to u_0$ with some constant $\mu$ not
depending on $t$. This mean that all the moments $c_n$ are defined
up to a nonzero constant $\mu$. Hence if condition \re{u0c0} holds
then we can always assume that $\kappa=1$, i.e. $u_0=c_0$.

There is a trivial generalization of \re{u0c0} leading again to the Schr\"odinger type of the Sturm-Liouville equation \re{schr_toda}. Indeed, it is sufficient to put $b_{-1}=\beta$, where $\beta$ is a constant not depending on $t$. Equivalently, this means 
\be
\frac{\dot c_0}{c_0} - \frac{\dot u_0}{u_0} =\beta \lab{ini_beta} \ee
Then equation \re{schr_toda} becomes the Schr\"odinger equation
\be \ddot \psi + (u_0(t) -(z+\beta)^2/4)
\psi =0 \lab{shcr_beta}. \ee
Condition \re{ini_beta} means that 
\be
c_0(t) = C e^{\beta t} u_0(t), \lab{c_0_beta} \ee
with and arbitrary constant $C$. Whence 
\be
b_0 = \frac{\dot c_0}{c_0} = \frac{\dot u_0}{u_0} + \beta \lab{b_0_beta}. \ee
This means that the coefficient $b_0$ is shifted by the constant $\beta$. By induction, it is easy to show that this is valid for all coefficients: if  $\{b_n(t), u_n(t)\}$ is unique solution of the Toda chain equations \re{Toda_eq} corresponding to the initial conditions $b_0=\frac{\dot u_0}{u_0}$ then $\{b_n(t)+\beta, u_n(t)\}$ is unique solution corresponding to the initial conditions $b_0=\frac{\dot u_0}{u_0}+\beta$. Thus the case $\beta \ne 0$ corresponds to a trivial shift of all coefficients $b_n$ by the same constant $\beta$. So, in what follows we can assume that $\beta=0$ (i.e. $b_{-1}=0$ ) without loss of generality.

\section{Boundary conditions of reflection type}
\setcounter{equation}{0}
In this section we consider restrictions for the Toda chain solutions $u_n(t), b_n(t)$ arising from
the boundary condition $b_{-1}(t)=0$. We have a simple
\begin{lem} \lab{refl}
Assume that the boundary condition $b_j(t)=0$ holds for some
fixed integer $j =0, \pm 1, \pm 2, \dots$. This condition is
equivalent to conditions \be u_{j+n} = u_{j-n+1}, \quad
b_{j+n-1}=-b_{j-n+1}, \quad n=0, \pm 1, \pm
2,\dots, \lab{sym_ub} \ee  
on solutions of the Toda chain.
\end{lem}

The proof of this Lemma is quite elementary. Indeed, assume that
$b_j=0$. From the first equation of \re{Toda_eq} we have
$u_{j+1}=u_j$. Then from the second equation of \re{Toda_eq}
(taken for $n=j,j+1$) we obtain $b_{j+1}=-b_{j-1}$. The statement
of the Lemma is obtained then by induction. The inverse statement
is trivial.

Solutions of the Toda chain with such "reflection" behavior are equivalent to the type B Toda lattice solutions introduced by Ueno and Takasaki \cite{UeTa}.

This boundary condition has a
simple mechanical meaning. Indeed, consider the Hamiltonian of the
Toda chain \be H= \sum_{k=N_1}^{N_2} p_k^2/2 +
\sum_{k=N_1}^{N_2-1} \exp(q_k-q_{k+1}) \lab{Ham_Toda}, \ee where
$p_k,q_k$ are standard canonical conjugated dynamical variables
with the Poisson brackets $\{q_k,q_l\} = \{p_k,p_l\} =0, \;
\{q_k,p_l\} = \delta_{kl}$. The limits $N_1, N_2$ may be finite or
infinite.

The "standard" boundary conditions for the Toda chain are chosen as follows.

Assume that $q_{-1}=-\infty, \: q_{N}=\infty$. Then we can put $N_1=0, \: N_2 = N-1$. In this case we deal with a finite
Toda chain consisting of $N$
particles $q_0, q_1, \dots, q_{N-1}$ (sometimes this model is called the "Toda molecule").

If $q_{-1}= -\infty$ (without other restrictions) then we have semi-infinite (or restricted) Toda chain.
This means that $N_1=0$ and $N_2=\infty$. Finally, if all $q_i$ are finite we have unrestricted (twicely
infinite) Toda chain with $N_1 = -\infty, \; N_2 = \infty$.

If one defines new variables $u_k, b_k$ by
\be
b_k = -p_k=-\dot q_k, \qquad u_k = \exp(q_{k-1}-q_{k}),
\lab{bu_qp} \ee
then we return to already considered standard Toda chain equations
\be
\dot b_k = u_k-u_{k-1}, \qquad \dot u_k = u_k (b_{k+1} - b_k).
\lab{Toda_eq_st} \ee
The Toda molecule boundary conditions are then equivalent to
\be
u_{0}(t)=u_{N}(t)=0.
\lab{ini_mol} \ee
For semi-infinite Toda chain we have the only boundary condition
\be
u_0(t)=0.
\lab{ini_semi} \ee

Due to translational invariance of the Toda chain we can conclude that the Toda molecule consisting of
$N$ particles can be realized if and only if the condition $u_{j} = u_{N+j}=0$ holds where $j$
is a fixed integer and $N$ is a fixed positive integer.

There are however "nonstandard" but still very natural boundary conditions. 

Fix $q_0(t) \equiv 0$. Then it is almost obvious from mechanical considerations that
the chain is completely anti-symmetric with respect to the point
$q_0=0$, i.e.\ \be q_{-n}(t) = - q_n(t), \quad p_{-n}(t) = - p_n(t).
\lab{sym_qp} \ee But conditions \re{sym_qp} are equivalent to
reflection conditions \re{sym_ub} for $j=0$. Due to translational
symmetry of the Toda chain we see that the boundary condition
\re{sym_ub} is equivalent to the condition $q_j(t) \equiv 0$ (i.e.
in the model described by Hamiltonian \re{Ham_Toda} we just fix
one of the particle unmoving).

On the other hand we see that boundary condition $b_{-1}(t) \equiv
0$ is equivalent to the choice $u_0(t)=c_0(t)$ when the
Sturm-Liouville equation \re{schr_toda} is reduced to a simple
Schr\"odinger equation \re{shcr}. Thus the "reflection" solutions of type B (in the sense of \cite{UeTa})
\re{sym_ub} of the Toda chain correspond to solutions of the
Schr\"odinger equation.

As a simple consequence of this boundary condition we have
\begin{pr} \lab{fin}
Assume that the boundary condition $b_{-1}=0$ is taken. Assume that for corresponding Toda chain solution a condition
$u_{N-1} \equiv 0$ holds for some positive $N=1,2,3,\dots$. Then necessarily $u_{-N} \equiv 0$.
\end{pr}
The proof follows immediately from formulas \re{sym_ub} for
$j=-1$. From this proposition it follows that the reflection
boundary condition $b_{-1}=0$ together with the restriction condition $u_{N-1}=0$
leads in fact to the Toda molecule. Indeed, we then have that
$u_{N-1}=u_{-N}=0$. This means that we deal with the Toda molecule
consisting of $2N-1$ particles. Note that in this case the total
number of particles is necessarily {\it odd}.

\section{Toda chain, moments  and KdV densities}
\setcounter{equation}{0}
The conserved densities $\sigma_m(x)$ of the KdV equation are
determined through the following differential-recurrence relations
\cite{Nov}: \be \sigma_{m+1}(x) =  \sigma'_m(x) + \sum_{k=1}^{m-1}
\sigma_k(x) \sigma_{m-k}(x), \quad m=1,2,\dots \lab{sigma_chain}, \ee
where initial condition is $\sigma_1(x) = -U(x), \; \sigma_2(x) =
-U'(x)$ and $U(x;t)$ satisfy the KdV equation
$$
U_{t} - 6UU_x + U_{xxx} = 0.
$$
The potential $U(x;t)$ is related with the Schr\"odinger equation
\be -\psi''(x;t) + U(x;t) \psi(x;t) = E \psi(x;t). \lab{kdv_schr}
\ee The well known Lax property of KdV states that under KdV
evolution the energy $E$ is a conserved quantity. All other
conserved quantities $I_m$ can be constructed as integrals from
the {\it odd} densities \be I_m[u]= \int \sigma_{2m-1}(U,U_x,
U_{xx}, \dots U_{m-2} )dx, \quad m=1,2,\dots \lab{I_m}. \ee Note
that $\sigma_{2m-1}$ are polynomials of the variable $U(x;t), U_{x}(x,t), U_{xx}(x,t), \dots$. 
The even functions $\sigma_{2m}$ are not important in theory of KdV \cite{Nov} because
they give complete derivatives (with respect to $x$) and hence
lead to only trivial integrals.

Now we can relate the KdV densities $\sigma_m(x)$ with the Toda
chain moments $c_n(t)$. We put $c_0(t) =u_0(t)= -U(t)$. Then it is elementary verified  that \be \sigma_m(t) = c_{m-1}(t), \quad m=1,2,3,\dots . 
\lab{sigma_c} \ee This means that system of the nonlinear equations  \re{sigma_chain} for the KdV densities coincides with
the system of equations \re{chain_mom} for the unrestricted Toda  under the additional condition \be
u_0(t) = c_0(t),  \lab{u0=c0} \ee which is equivalent to the
condition $b_{-1}(t) =0$. Thus the theory of conserved densities of
the KdV equation can be reduced to 
the theory of the unrestricted Toda chain with the additional
condition \re{u0=c0}.

Note that due to the condition $\sigma_1(x) = -U(x)$ we see that
the Schr\"odinger potential $U(x)$ coincides with the
Toda "potential" $u_0(t)$: \be U(x) = - u_0(x) \lab{Uu}. \ee This
observation has several possible applications.

First of all, we can apply already developed the Toda chain
analysis to the theory of KdV (and Schr\"odinger) solutions. In
particular, we can relate these solutions with the theory of
corresponding orthogonal polynomials.

Second, starting from exactly solvable quantum mechanical
potentials, we can construct corresponding Toda chain solutions
and corresponding orthogonal polynomials.

\section{Rational Stieltjes function and reflectionless potentials}
\setcounter{equation}{0}
Assuming the condition $u_0(t) = c_0(t)$ (or equivalently $b_{-1}(t) =0$), we see that
\be F(z;t) = \frac{\dot
\psi(z;t)}{\psi(z;t)} + z/2 \lab{F_psi_S}. \ee

Let us consider the case of the rational Stieltjes function $F(z;t)$, i.e.\
$$
F(z;t)=\frac{Q_1(z;t)}{Q_2(z;t)},
$$
where $Q_1(z;t)$ and $Q_2(z;t)$ are polynomials in $z$ with coefficients depending on $t$.
Without loss of generality we can assume that $Q_2(z)$ is a monic polynomial of a degree $N$: $Q_2(z) = z^N + O(z^{N-1})$.
From definition \re{St_F} it follows that $\deg(Q_1(z)) = N-1$ while from the Riccati equation
\re{Ric_F1} it follows that all zeros of the polynomial $Q_2(z)$ are simple:
\be
Q_2(z;t) = (z-a_1(t))(z-a_2(t))\dots (z-a_N(t)), \lab{Q_2} \ee
{i.e.\ }the functions $a_k(t)$ are simple zeros of the polynomial $Q_2(z;t)$.
Then we can present $F(z;t)$ in an equivalent form as
\be
F(z;t) = \sum_{k=1}^N \frac{A_k(t)}{z-a_k(t)} \lab{F_rat} \ee
with some functions $A_k(t)$ satisfying the condition
\be
\sum_{k=1}^N A_k(t) = c_0(t). \lab{sum_A} \ee
Condition \re{sum_A} follows from definition \re{St_F}.

Substituting expression \re{F_rat} into Riccati equation \re{Ric_F1} and assuming $u_0=c_0$
we obtain immediately the condition
\be
A_k(t) = - \dot a_k(t). \lab{A_a} \ee
For poles $a_k(t)$ we obtain from \re{Ric_F1} a system of $N$ nonlinear differential equations
\be
\ddot a_k = a_k \dot a_k + 2 \sum_{m \ne k} {\frac{\dot a_k \dot a_m}{a_k-a_m}}, \quad k=1,2,\dots, N. \lab{eqs_a} \ee
It was shown in \cite{VDP} that these equations describe an integrable rational Ruijsenaars-
Schneider particle system with harmonic term. We thus see that this system is equivalent to the Toda molecule with additional boundary condition $b_{-1}(t) \equiv 0$.

From \re{A_a} and \re{F_psi_S} it follows that the function $\psi(z;t)$ can be presented as
\be
\psi(z;t) = e^{-zt/2}(z-a_1(t)) \dots (z-a_N(t)) = e^{-zt/2} Q_2(z;t), \lab{psi_Q} \ee
{i.e.\ }that the wave function is a polynomial in $z$ multiplied by the exponential function $e^{-zt/2}$ corresponding
to the "free motion" (when $u_0(t) =0$).

It is well known (see, e.g. \cite{VDP}) that all such solutions of the Schr\"odinger equation \re{shcr} are in one-to-one correspondence with the so-called reflectionless  potentials (sometimes called the Bargmann potentials \cite{Barg}) obtained from the free Schr\"odinger equation with $u_0(t)=0$ by application of $N$ succeeding Darboux transforms. We thus see that all rational solutions of the Riccati equation for the Stieltjes function correspond to the reflectionless potentials of the Schr\"odinger equation (and vice versa). Note that the system of nonlinear differential equations \re{eqs_a} appeared also in \cite{Leznov} n order to give an effective description of the Bargmann potentials.

Equations \re{eqs_a} are completely integrable, i.e. there exists $N$ independent integrals of motion. This was shown in \cite{VDP} where all these integrals were derived explicitly.  In \cite{Leznov} it was also noticed that these integrals can be presented in the compact form
\be
\dot a_k =\frac{V(a_k^2)}{\Omega'(a_k^2)}, \lab{int_Leznov} \ee 
where $\Omega(x)=(x-a_1^2)(x-a_2) \dots(x-a_N^2)$ and $V(x) = (x-\mu_1)(x-\mu_2)\dots (x-\mu_N)$. The parameters $\mu_1, \mu_2, \dots \mu_N$ are arbitrary and they play the role of the integrals of motion.

Equations \re{int_Leznov} look like the Dubrovin equations \cite{Dubrovin} in the theory of finite-gap potentials. This is not surprising because in \cite{PSZ} it was shown that the Dubrovin equations describe time dynamics of the Toda chain solutions corresponding to second degree forms (finite-gap solutions). The solutions \re{F_rat} correspond to a degeneration of the finite-gap solutions of the Toda chain.

There are simple consequences following from the choice of $F(z;t)$ as a rational function.
\begin{pr}
If the Stieltjes function $F(z;t)$ for orthogonal polynomials $P_n(x;t)$ is a rational function \re{F_rat} then

(i) The polynomials $P_n(x;t)$ are orthogonal on the finite set of points $a_k$:
\be
\sum_{k=1}^N A_k(t) P_n(a_k(t);t)P_m(a_k(t);t) = h_n(t) \delta_{nm} \lab{fin_ort} \ee
with concentrated masses $A_k(t) = -\dot a_k(t)$.

(ii) the monic orthogonal polynomial $P_N(x;t)$ has the explicit expression
\be
P_N(x;t) = (x-a_1(t)) (x-a_2(t)) \dots (x-a_N(t)), \lab{P_N_exp} \ee

(iii) the moments $c_n(t)$ have the explicit expression
\be
c_n(t) = \sum_{k=1}^N A_k a_k^n = -\sum_{k=1}^N \dot a_k a_k^n, \lab{moms_A} \ee

(iv) the moments $c_n(t)$ satisfy the recurrence relation
\be
\sum_{k=0}^N B_k(t) c_{n+k}(t) =0, \quad n=0,1,2, \dots, \lab{mom_fin} \ee
where the coefficients $B_0(t), B_2(t), \dots B_N(t)$ do not depend on $n$.

(v) the Hankel determinant $H_{N+1}(t)$ vanishes $H_{N+1}(t) \equiv 0$, whereas the Hankel determinant $H_N(t)$ is proportional to square of the Vandermond determinant from parameters $a_1, a_2, \dots, a_N$:
\be
H_N(t) = A_1(t) A_2(t) \dots A_N(t) \: \prod_{i<k}(a_i(t)-a_k(t))^2. \lab{vander} \ee

\end{pr}

Proofs of statements (i)-(iii) follows easily from theory polynomials orthogonal on a finite set of points (see, e.g. \cite{Atk}, \cite{Chi}).
Statement (iv) follows directly from (iii); moreover it follows from the well known theorem about rational generating functions \cite{Lando}. Statement (v) follows from (iii) after simple manipulations with corresponding determinants.

There is simple matrix interpretation of the above equations for the quantities $a_i(t)$. Indeed, let us introduce the tridiagonal (Jacobi) matrix $J$ of size $N \times N$
$$
J=\left[
\begin{array}{ccccc}
b_0 & 1 & 0 & \cdots & 0 \\
u_1 & b_1 & 1 &  & \vdots  \\
0 & \ddots & \ddots & \ddots & 0 \\
\vdots & & u_{N-2} & b_{N-2} & 1 \\
0 & \cdots & 0 & u_{N-1} & b_{N-1}
\end{array}
\right].
$$
Introduce also the matrix $A$ which is the lower-triangular part of the matrix $J$, i.e.
$$
A=\left[
\begin{array}{ccccc}
0 & 0 & 0 & \cdots & 0 \\
u_1 & 0 & 0 &  & \vdots  \\
0 & \ddots & \ddots & \ddots & 0 \\
\vdots & & u_{N-2} & 0 & 0 \\
0 & \cdots & 0 & u_{N-1} & 0
\end{array}
\right].
$$

If we assume that the coefficients $b_0(t), b_1(t), \dots b_{N-1}(t)$ and $u_1(t), u_2(t), \dots u_{N-1}(t)$ satisfy the Toda chain equations \re{Toda_eq} then the quantities $a_i(t), \; i=1,2,\dots, N$ are simple eigenvalues of the Jacobi matrix $J(t)$.  

In matrix form equations \re{Toda_eq} can be presented as
\be
\dot J = [J,A] - u_0 M, \lab{J_comm} \ee
where $[A,J]$ stands for commutator of two matrices and $M$ is the matrix with the only nonzero entry $M_{00}=1$. Note that algebraic relation \re{J_comm} is a perturbation of the well known Lax pair relation $\dot J = [J,A]$ with the additional term $-u_0 M$. If $u_0=0$ then we have the standard restricted Toda molecule and all eigenvalues $\lambda_i$ of the Jacobi matrix $J(t)$  are the integrals of motion: $\dot \lambda_i =0$. However for $u_0 \ne 0$ equation \re{J_comm} is NOT in the Lax form and hence the matrix $J(t)$ is no more isospectral. This means that the eigenvalues $a_i(t)$ do depend on $t$.

\section{Solutions of type B and C and the spectral problem for tridiagonal per-skew symmetric matrices}
\setcounter{equation}{0}
We have already identified the Schr\"odinegr-type solutions with the solutions of type B proposed by Ueno and Takasaki  \cite{UeTa}. In this section we consider solutions of type B and C from the point of view of spectral theory of corresponding Jacobi matrices.   

The type B solutions correspond to the boundary condition $b_{-1}=0$ which is equivalent to the reflection conditions 
\be
u_n = u_{-1-n}, \quad b_{n}=-b_{-2-n}. \lab{ref_B} \ee
The type C solutions correspond to the boundary condition $b_{-1}=-b_0$ which is equivalent to the reflection conditions
\be
u_n=u_{-n}, \quad b_n=-b_{-n-1}. \lab{ref_C} \ee
It is clear that the solutions of the type B (i.e. $b_{-1}=0$) coincide with already considered special solutions of the Toda chain corresponding to the pure Schr\"odinger equation. Solutions of the type C do not correspond to the Schr\"odinger equation. In this case equation \re{schr_toda} becomes 
\be \ddot \psi + (u_0 +\dot b_{0}/2 -(b_{0}+z)^2/4) \psi =0.
\lab{schr_toda_C} \ee
It corresponds to quadratic pencil eigenvalue problems, i.e. 
\be
(K+ \lambda L + \lambda^2 M)\psi =0, \lab{qua_pen} \ee
with 3 operators $K,L,M$.

Consider the finite-dimensional case of solutions of types B and C. This means the boundary condition $u_N=0$ for some $N=1,2,\dots$. From the reflection conditions it follows that $u_{-1-N}=0$ for the type B and $u_{-N}=0$ for the type C. This leads to finite-dimensional solutions of the Toda molecule type. 

For the type B let us introduce the tridiagonal matrices of size $2N+1$
\ba
J=\left[
\begin{array}{ccccc}
b_{-N-1} & 1 & 0 & \cdots & 0 \\
u_{-N} & b_{-N} & 1 &  & \vdots  \\
0 & \ddots & \ddots & \ddots & 0 \\
\vdots & & u_{N-2} & b_{N-2} & 1 \\
0 & \cdots & 0 & u_{N-1} & b_{N-1}
\end{array}
\right]. \lab{J_type_B}
\ea
Due to conditions \re{ref_B} this matrix has a specific symmetry structure
\ba
J=\left[
\begin{array}{ccccc}
-b_{N-1} & 1 & 0 & \cdots & 0 \\
u_{N-1} & -b_{N-2} & 1 &  & \vdots  \\
0 & \ddots & \ddots & \ddots & 0 \\
\vdots & & u_{N-2} & b_{N-2} & 1 \\
0 & \cdots & 0 & u_{N-1} & b_{N-1}
\end{array}
\right]. \lab{J_type_B_r}
\ea
Similarly, for the type C we can introduce the tridiagonal matrix of size $2N$
\ba
J=\left[
\begin{array}{ccccc}
b_{-N} & 1 & 0 & \cdots & 0 \\
u_{-N+1} & b_{-N+1} & 1 &  & \vdots  \\
0 & \ddots & \ddots & \ddots & 0 \\
\vdots & & u_{N-2} & b_{N-2} & 1 \\
0 & \cdots & 0 & u_{N-1} & b_{N-1}
\end{array}
\right]. \lab{J_type_C}
\ea
Again, due to conditions \re{ref_C} we have a specific symmetry structure 
\ba
J=\left[
\begin{array}{ccccc}
-b_{N-1} & 1 & 0 & \cdots & 0 \\
u_{N-1} & -b_{N-2} & 1 &  & \vdots  \\
0 & \ddots & \ddots & \ddots & 0 \\
\vdots & & u_{N-2} & b_{N-2} & 1 \\
0 & \cdots & 0 & u_{N-1} & b_{N-1}
\end{array}
\right]. \lab{J_type_C_r}
\ea
In order to clarify symmetry properties of the matrices \re{J_type_B_r} and \re{J_type_B_r} we introduce the reflection ( exchange) matrix
\ba
R=\left[
\begin{array}{ccccc}
0 & 0 &  \cdots & 0 &  1 \\
0 & 0 &  \cdots & 1  & 0 \\
\ddots & \ddots & \ddots & \ddots & \ddots \\
0 & 1 & \cdots & 0 & 0 \\
1 & 0 & \cdots & 0 & 0
\end{array}
\right]. \lab{R}
\ea
The matrix $A$ is called the {\it persymmetric} if it is symmetric with respect to the reflection $R$:
\be
AR = R A^{T}, \lab{A_per} \ee 
where $A^{T}$ means transposed matrix. Similarly, the matrix $A$ is called the {\it per-skew} symmetric if
\be
AR = -R A^{T}. \lab{A_skew} \ee  
In particular, the persymmetric tridiagonal matrix looks as 
\ba
A=\left[
\begin{array}{ccccc}
b_{0} & 1 & 0 & \cdots & 0 \\
u_{1} & b_{1} & 1 &  & \vdots  \\
0 & \ddots & \ddots & \ddots & 0 \\
\vdots & 0 & u_{2} & b_{1} & 1 \\
0 & \cdots & 0 & u_{1} & b_{0}
\end{array}
\right], \lab{per_tri}
\ea
while tridiagonal per-skew symmetric matrix looks as 
\ba
A=\left[
\begin{array}{ccccc}
b_{0} & 1 & 0 & \cdots & 0 \\
u_{1} & b_{1} & 1 &  & \vdots  \\
0 & \ddots & \ddots & \ddots & 0 \\
\vdots & 0 & -u_{2} & -b_{1} & -1 \\
0 & \cdots & 0 & 0-u_{1} & -b_{0}
\end{array}
\right]. \lab{skew_tri}
\ea
It is seen that the matrices \re{J_type_B} and \re{J_type_C} look very similar to per-skew symmetric tridiagonal matrices. In fact, they differ from per-skew symmetric tridiagonal matrices by a trivial similarity transformation.

Indeed, let us introduce the diagonal matrix $S$ with the entries
\be
S_{ik} = (-1)^i \: \delta_{ik}. \lab{S_matr} \ee 
Obviously, $S$ is an involution, i.e. $S^2=I$, where $I$ is the identical matrix.
For any tridiagonal matrix 
\ba
J=\left[
\begin{array}{ccccc}
b_{0} & 1 & 0 & \cdots & 0 \\
u_{1} & b_{1} & 1 &  & \vdots  \\
0 & \ddots & \ddots & \ddots & 0 \\
\vdots & & u_{N-2} & b_{N-2} & 1 \\
0 & \cdots & 0 & u_{N-1} & b_{N-1}
\end{array}
\right], \lab{J_arb} \ea
we have
\ba
SJS=\left[
\begin{array}{ccccc}
b_{0} & -1 & 0 & \cdots & 0 \\
-u_{1} & b_{1} & -1 &  & \vdots  \\
0 & \ddots & \ddots & \ddots & 0 \\
\vdots & & -u_{N-2} & b_{N-2} & -1 \\
0 & \cdots & 0 & -u_{N-1} & b_{N-1}
\end{array},
\right], \lab{SJS} \ea
{i.e.\ }under the transformation $S$ the off-diagonal entries change their sign while diagonal entries remain the same. 

Whence,we have the 
\begin{pr}
The Jacobi matrices corresponding to the types B and C of the Toda chain satisfy the defining relation
\be
SR J RS = - J^T. \lab{SRJ} \ee
They are similar to per-skew symmetric tridiagonal matrices. The type B corresponds to the matrices with the odd size $2N+1$ while the type C corresponds to the matrices with the even size $2N$. 
\end{pr}
Thus the spectral properties of the matrices of type B and C coincide with the spectral properties of per-skew symmetric matrices.

Spectral theory of persymmetric tridiagonal matrices is well developed (see, e.g. \cite{BoGo} where the algorithm for the inverse spectral problem is proposed).  Spectral properties of skew-persymmetric matrices are discussed in \cite{Trench}. Here we present the main spectral properties of the per-skew symmetric matrices.

With any tridiagonal matrix \re{J_arb} one can associate a system of orthogonal polynomials $P_n(x)$ defined by the three-term recurrence relation
\be
P_{n+1} + b_n P_n(x) + u_n P_{n-1}(x) = x P_n(x), \quad n=0,1,\dots, N-1 \lab{rec_P}, \ee
and initial conditions
\be
P_{-1}=0, \; P_0(x)=1. \lab{ini_P} \ee
In case if all off-diagonal entries are positive $u_i>0$, the spectrum of the Jacobi matrix is simple
\be
J \chi^{(s)} = x_s \chi^{(s)}, \lab{eigen_chi} \ee  
where $\chi^{(0)}, \chi^{(1)}, \dots, \chi^{(N-1)}$ are linearly independent eigenvectors corresponding to the eigenvalues $x_s$. These eigenvectors can be presented in terms of orthogonal polynomials
\be
\chi^{(s)} =\{P_0(x_s), P_1(x_s), \dots, P_{N-1}(x_s)\}. \lab{chi_P} \ee
The eigenvalues $x_s$ are distinct zeros of the "final" polynomial $P_N(x)$:
\be
P_N(x) = (x-x_0)(x-x_1)\dots (x-x_{N-1}). \lab{PN} \ee
The polynomials $P_n(x)$ are orthogonal with respect to a discrete measure on the real axis
\be
\sum_{s=0}^{N-1} P_n(x_s) P_m(x_s) w_s = h_n \: \delta_{nm}, \lab{ort_P_s} \ee
where $h_n=u_1 u_2 \dots u_n$ is the normalization factor. The discrete weights are positive $w_s>0$; they can be uniquely determined from the matrix $J$.

From general theory of the per-skew symmetric matrices \cite{Trench} it is easy to derive the
\begin{pr}
The eigenvalues $x_s$ of the positive definite per-skew symmetric Jacobi matrix are symmetric with respect to zero, i.e.
\be
x_{N-s-1}=-x_s, \quad s=0,1,\dots,N. \lab{sym_x} \ee
\end{pr}
In particular, if the dimension $N$ of the matrix is odd then one of these roots is zero: $x_{(N-1)/2}=0$. If $N$ is even then we have $N/2$ distinct positive zeros and corresponding negative zeros with the same absolute values. Define the characteristic polynomial of these roots:
\be
\Omega(x) = (x-x_0)(x-x_1) \dots (x-x_{N-1}). \lab{Omega} \ee 
If $N$ is even then the polynomial $\Omega(x)$ is even:
\be
\Omega(x) = (x^2-x_0^2)(x^2-x_1^2) \dots (x^2 - x_{N/2-1}^2). \lab{Om_even} \ee
If $N$ is odd then $\Omega(x)$ is odd:
\be
\Omega(x) = x (x^2-x_0^2)(x^2-x_1^2) \dots (x^2 - x_{(N-3)/2}^2). \lab{Om_odd} \ee 
Note that $\Omega(x)=P_N(x)$ which follows from the fact that the roots of the polynomial $P_N(x)$ coincide with the eigenvalues $x_s$.

It is convenient to introduce the orthonormal polynomials $\pi_n(x)$ by the formula 
\be
\pi_n(x) = \frac{P_n(x)}{\sqrt{h_n}}. \lab{orto_p} \ee
These polynomials satisfy the recurrence relation
\be
a_{n+1} \pi_{n+1}(x) + b_n \pi_n(x) + a_n \pi_m(x) = x \pi_n(x), \lab{orto_rec} \ee
with $a_n= \sqrt{u_n}>0$. Orthogonality relation for the polynomials $\pi_n(x)$ reads
\be
\sum_{s=0}^{N-1} \pi_n(x_s) \pi_m(x_s) w_s =  \delta_{nm}. \lab{orto_w} \ee

\begin{pr}
Assume that $J$ is a per-skew symmetric Jacobi matrix with positive off-diagonal entries $u_i$. Then:

{\normalfont (i)} the weights $w_s$ satisfy the properties
\be
w_{s} w_{N-s-1} = \frac{h_{N-1}}{{{\Omega'}^2(x_s)}},  \lab{ww} \ee

{\normalfont (ii)} the orthonormal polynomials $\pi_{N-1}(x)$ satisfy the property
\be
\pi_{N-1}(x_s) \pi_{N-1}(x_{N-s-1}) = 1 , \quad s=0,1,\dots, N-1. \lab{PPh} \ee

Moreover, if the zeros $x_s$ satisfy the symmetry condition {\normalfont \re{sym_x}} then any of the properties {\normalfont (i)} or {\normalfont (ii)} determine the per-skew symmetric Jacobi matrix $J$. 
\end{pr} 
The proof of this proposition follows from general properties of orthogonal polynomials corresponding to the Jacobi matrices $J$ and $J^* = R A^T R$ (see \cite{VZ1}, \cite{VZ_PST} for details).

Note that in the special case of pure persymmetric matrix ({i.e.\ }all diagonal entries $b_i$ vanish) we have the condition \cite{VZ_PST}
\be
\pi_{N-1}(x_s) =(-1)^{N-s-1}, \lab{pi_per} \ee 
where it is assumed that the eigenvalue are ordered by increase: $x_0<x_1<\dots < x_{N-1}$. In this case the polynomial $\pi_{N-1}(x)$ can be restored uniquely by the Lagrange interpolation formula. We thus know explicitly two monic polynomials: $\Omega(x)=P_N(x)$ and $P_{N-1}(x)$. This gives an efficient algorithm to restore uniquely the persymmetric Jacobi matrix $J$ \cite{VZ_PST}.

In case of per-skew symmetric matrices we can put 
\be
\pi_{N-1}(x_s) = (-1)^{s+1} \tau_s, \quad s=0,1,\dots, N/2-1,  \quad  \quad N \quad \mbox{even} \lab{x_s_even} \ee
and 
\be
\pi_{N-1}(x_s) = (-1)^{s} \tau_s, \quad s=0,1,\dots, (N-1)/2-1, \; x_0 =(-1)^{N(N-1)/2}, \quad  N \quad \mbox{odd} \lab{x_s_odd} \ee
with arbitrary positive parameters $\tau_s$. Then all values $\pi_{N-1}(x_s)$ can be determined by \re{PPh}. Again, we know explicitly the polynomials $P_N(x)$ and $P_{N-1}(x)$ and the per-skew symmetric Jacobi matrix $J$ can be restored uniquely using the same algorithm as in \cite{VZ_PST}.

Alternatively, one can start with the prescribed discrete weights. Assume e.g. that $N$ is even. We can take $\rho_0, \rho_1, \dots, \rho_{N/2-1}$ as arbitrary positive parameters. Define then $\rho_i = 1/\rho_{N-1-i}$ for $i=N/2, N/2+1, \dots, N-1$. We can identify 
\be
w_i = \mu \rho_i, \quad i=0,1,\dots,N-1, \lab{w_rho} \ee 
where the normalization coefficient $\mu$ can be found from the condition $w_0 + w_1 + \dots + w_{N-1}=1$. Starting with these data,we can construct uniquely the polynomials $P_n(x)$ and corresponding per-skew symmetric Jacobi matrix $J$ using standard algorithms  \cite{BoGo}.

If all diagonal entries vanish, i.e. $b_n=0$, then the per-skew symmetric matrix becomes the ordinary persymmetric matrix. In this case it is sufficient to start with prescribed eigenvalues $x_s$ satisfying the symmetry condition \re{sym_x}. The corresponding persymmetric Jacobi matrix $J$ can be restored uniquely \cite{BoGo}. We thus see that in contrast to the case of the persymmetric matrices, the inverse spectral problem for per-skew symmetric matrices needs more information than knowledge of the eigenvalues only.

\section{Simple examples of the Toda-Schr\"odinger correspondence}
\setcounter{equation}{0}
Assume that a potential $u_0(t)$ is chosen as the initial condition.
Then by \re{u0=c0} the recurrence coefficient $b_0(t)$ is uniquely expressible via the "potential" $u_0(t)$: $b_0(t)=\dot c_0/c_0= \dot
u_0/u_0.$ Clearly, the next recurrence coefficients $u_n(t), b_n(t)$
are expressible uniquely in terms of $u_0(t)$.

Consider several simple examples.

Let us choose
$$
u_0(t) = \alpha/t^2,
$$
with an arbitrary parameter $\alpha$ (this corresponds to the simplest quantum mechanical centrifugal potential).
Then it is easily verified that
$$
u_n(t) = \frac{n(n+1) + \alpha}{t^2}, \quad b_n(t) = -\frac{2(n+1)}{t}.
$$
In this example we obtain that orthogonal polynomials $P_n(x;t)$ coincide with the associated Laguerre polynomials \cite{PSZ}.

Quite similarly, choosing
$$
u_0(t) = \frac{\alpha}{\cos^2(t)},
$$
one obtains the solution
$$
u_n(t) = \frac{\alpha + n(n+1)}{\cos^2(t)}, \quad b_n = 2(n+1) \: \tan(t).
$$
These recurrence coefficients correspond to the associated Meixner-Pollaczek polynomials \cite{PSZ}. 

Finally, consider the choice
\be
u_0(t) = \frac{N(N+1)}{\cosh^2(t)}
\lab{u_sol} \ee
(the N-solitonic potential). Then we obtain
\be
u_n(t) = \frac{N(N+1)-n(n+1)}{\cosh^2(t)}, \quad b_n = -2(n+1) \: \tanh(t).
\lab{ass_Kr} \ee
The recurrence coefficients \re{ass_Kr} correspond to the associated Krawtchouk polynomials \cite{KLS}. Moreover, it is seen that $u_N=0$ and hence we have a special case when the Stieltjes function is rational. Let us consider this case in more details. 

From results of the previous section it follows that the solution of the Schr\"odinger equation   \re{shcr} with the potential \re{u_sol} can be presented as
\be
\psi_N(t;z) = e^{-tz/2} Q_N(z;t), \lab{sol_psiN} \ee
where $Q_N(z;t)$ is a monic polynomial of the $n$-th degree
\be
Q_N(z;t) =z^N + r_{N-1}(t) z^{N-1} + \dots + r_0(t) \lab{sol_QN}, \ee
with the coefficients $r_k(t)$ depending on $t$. Let us stress that solution \re{sol_psiN} is NOT the general solution of the Schr\"odinger equation \re{shcr}. It is the unique special solution which satisfies the asymptotic condition
$$
F(z;t) = \frac{\dot \psi}{\psi} + \frac{z}{2} =  c_0(t) z^{-1} + O(z^{-2}).  
$$
The polynomial $Q_N(z;t)$ can be constructed recursively, using the Darboux transformation of the Schr\"odinger equation.

Recall basic facts concerning the Darboux transform for the Schr\"odinger equation (see. e.g. \cite{MS}). Let $\psi(t)$ be a generic solution of the Schr\"odinger equation \re{shcr}.  Assume that the function $\phi(t)$ is a special solution of the same Schr\"odinger equation
\be
\ddot \phi(t) + (u_0(t) -\mu^2/4) \phi(t) =0, \lab{phi_eq} \ee 
with the spectral parameter $z$ equal to $\mu$. Then the function
\be
\tilde \psi(t) = \kappa \left(\dot \psi - \frac{\dot \phi}{\phi}  \psi \right) \lab{D_psi} \ee
is the generic solution of the Schr\"odinger equation
\be
{\ddot {\tilde \psi}} + (\tilde u_0(t) - z^2/4) {\tilde \psi},   \lab{D_SCH} \ee
where
\be
\tilde u_0(t) = u_0(t) + 2 \frac{d^2 \log \phi(t)}{dt^2}. \lab{uu} \ee
Note that $\kappa$ can be an arbitrary constant.
 
For the potential $u_0(t) =N(N+1) \cosh^{-2}(t)$ it is verified that the function 
\be
\phi_N(t) = \cosh^{N+1}(t) \lab{phi_N} \ee
is the desired special solution corresponding to the eigenvalue $\mu= 2(N+1)$. Then the Darboux transformation leads to the potential $u_0(t) = (N+1)(N+2) \cosh^{-2}(t)$, i.e. it is equivalent to the shift $N \to N+1$. The solution \re{sol_psiN} becomes
\be
\psi_{N+1}(t;z) = 2 \left(\psi_N(t;z) - \frac{\phi_N(t))}{\phi_N(t)} \psi_N(t;z) \right) =  e^{-tz/2} Q_{N+1}(z;t), \lab{D_N} \ee
where the polynomial $Q_{N+1}(z;t)$ is related with $Q_N(z;t)$ as
\be
Q_{N+1}(z;t) = \left(z-2(N+1)\tanh t \right)  Q_N(z;t) - 2 \dot Q_N(z;t). \lab{QQ} \ee 
From \re{QQ} it is seen that $Q_{N+1}(z;t)$ is a monic polynomial of degree $N+1$:
\be
Q_{N+1}(z;t) = z^{N+1} + O(z^n) \lab{Q_{N+1}}. \ee
Hence formula \re{D_N} gives the unique solution of the Schr\"odinger equation \re{shcr} with the potential $u_0(t)=(N+1)(N+2)\cosh^{-2}(t)$.

Clearly, $Q_0(z;t)=1$. Then all next polynomials $Q_1(z;t), Q_2(z;t), \dots$ are determined uniquely from relation \re{QQ}. It is easy to see that $Q_n(z,t)$ is also a polynomial of degree $N$ with respect to the variable $y=\tanh(t)$. Hence, relation \re{QQ} can be rewritten in the form
\be
Q_{N+1}(z;y) = (z  + 2(N+1) y) Q_N(z;y) -2 (1-y^2) \partial_y Q_N(z;y). \lab{QQ_y} \ee
Relation \re{QQ_y} is a special example of a class of relations for polynomials $Q_N(y)$ in the variable $y$ of degree $N$:  
\be
Q_{N+1}(y) = \tau(y) Q_N(y) + \sigma(y) \partial_y Q_N(y), \quad Q_0 =1, \lab{QQ_St} \ee
where $\tau(y)$ and $\sigma(y)$ are polynomial of degrees at most one and two.  These relations go back to Stieltjes. Their role for solutions of the Toda chain was considered in \cite{NaZhe} and \cite{ViZhe}. 

The first three polynomials $Q_N(z;y)$ are
\ba
&&Q_1(z;y) = z+2y, \; Q_2(z;y) = z^2 + 6yz + 4(3 y^2-1), \nonumber \\
&&Q_3(z;y)= {z}^{3}+12\,y{z}^{2}+ \left( -16+60\,{y}^{2} \right) z+24\,y \left( -3+5\,{y}^{2} \right), \lab{Q_123} \ea
where $y=\tanh(t)$.

The polynomial $Q_N(z;t)$ has simple zeros $a_i(t)$ satisfying non-linear equations \re{eqs_a}. For generic $t$ it is impossible to give explicit expressions for the functions $a_i(t)$. However, for $t=0$ (i.e. for $y=0$) the polynomials $Q_N(z;0)$ have explicit zeros: if $N=2j$ is even then
\be 
a_k(0) = \pm (2+4k), \quad k=0,1,\dots, j-1. \lab{a0_even} \ee
If $N=2j+1$ is odd then
\be 
a_k(0) = \pm (4k), \quad k=0,1,\dots, j. \lab{a0_odd} \ee
Formulas \re{a0_even} and \re{a0_odd} can be derived using the theory of classical orthogonal polynomials. Indeed, for $t=0$ the recurrence relation for the polynomials $P_n(x;0)$ has the form
\be
P_{n+1}(x;0) + (N-n)(N+n+1)P_{n-1}(x;0) = x P_n(x;0). \lab{rec_Hahn} \ee
This recurrence relation can be identified with a special class of the Hahn polynomials.

Recall that the monic Hahn polynomials $H_n(x;\alpha,\beta,M)$ depend on 3 parameters $\alpha,\beta,M$ and satisfy the recurrence relation \cite{KLS}
\be
H_{n+1}(x) + b_n H_n(x) + u_n H_{n-1}(x) = xH_n(x), \lab{rec_H} \ee 
with 
\be
b_n = A_n + C_n, \quad u_n = A_{n-1} C_n, \lab{bu_AC} \ee
where
\ba
&&A_n = \frac{(n + \alpha+ \beta + 1)(n + \alpha+ 1)(M- n)}{(2n+\alpha+\beta+1)(2n+\alpha+\beta+2)}, \nonumber \\
&&C_n = \frac{(n + \alpha+ \beta + M+ 1)(n + \beta)(M- n)}{(2n+\alpha+\beta+1)(2n+\alpha+\beta)}. \lab{AC_H} \ea
When $M$ is a positive integer, the Jacobi matrix $J$ corresponding to the Hahn polynomials, has the spectrum 
\be
x_s = s, \quad s=0,1,\dots, M. \lab{x_Hahn} \ee 
Consider the special case of the Hahn polynomials with
\be
\alpha=\beta=1/2, \quad M=N-1. \lab{spec_par} \ee
Under these conditions we have 
\be
u_n = \frac{(N+n+1)(N-n)}{16}, \quad b_n = \frac{N-1}{2}. \lab{spec_ub_H} \ee
Comparing recurrence coefficients \re{spec_ub_H} with \re{rec_Hahn} we conclude that the polynomials $P_n(x;0)$ coincide (up to a trivial affine transformation of the argument $x$) with the Hahn polynomials $H_n(1/2,1/2,N-1)$. This leads to the spectrum $x_s$ coinciding with \re{a0_even} and \re{a0_odd}.

We thus obtained self-similar solutions of simple form. They correspond to
solutions with separated variables of the Toda chain  (for details see, e.g. \cite{NaZhe} and \cite{PSZ}).

However the above examples with the elementary solutions for $u_n(t), b_n(t)$ are rather exceptional. In the next section we consider a less elementary example leading to so-called Vorob'ev-Yablonskii polynomials in the theory of the Painlev\'e-II equation.

\section{Linear potential and the Vorob'ev-Yablonskii polynomials}
\setcounter{equation}{0}
Recall that the Painlev\'e-II equation has the form \cite{JKM}
\be
\frac{d^2 V(t)}{dt^2} = 2 V^3(t) - 4tV(t) + 4(\alpha+1/2), \lab{P_II} \ee
with an arbitrary parameter $\alpha$. When $\alpha=N+1/2$ (and only in this case) with an arbitrary non-negative integer $N$, the unique rational solutions of the  Painlev\'e-II equation arise. These solutions have the form 
\be
V_N(t) = \frac{d}{dt} \: \log {\frac{H_{N+1}(t)}{H_N(t)}},  \lab{phi_H} \ee
where $H_n(t)$ are the Hankel determinants constructed from the moments
$$
H_n(t)=\det ||a_{i+k}(t)||_{i,k=0}^{n-1}.
$$
The moments $a_n(t)$ are connected by conditions \cite{JKM} \be a_{n+1}(t)
= \dot a_n + \sum_{s=0}^{n-1} a_s a_{n-1-s} \lab{P_2_chain} \ee
and initial conditions \be a_0(t)=t, \; a_1(t)=1 \lab{ini_P_2}. \ee
It is easily seen that equations \re{P_2_chain} coincide with equations \re{chain_mom} under identification $a_n(t)=c_n(t)$ and initial conditions
\be u_0(t) =c_0(t)=t. \lab{u0=t} \ee
(Note that this choice of $u_0(t)$ corresponds to the linear potential of the
Schr\"odinger equation having explicit solutions in terms of the Airy functions \cite{LL}). 
It is obvious from the Toda chain equations \re{Toda_eq} and initial conditions \re{u0=t} that both $u_n(t)$ and $b_n(t)$ are rational functions in $t$. Expressions of these rational functions become of more and more complicated
when $n$ increases. Moreover, from correspondence between
Toda chain and orthogonal polynomials \cite{PSZ} we have \be b_n(t) = \frac{\dot
h_n(t)}{h_n(t)} \lab{bh}, \ee where
$$
h_n(t) = \frac{H_{n+1}(t)}{H_n(t)},
$$
is the normalization coefficient of the orthogonal polynomials
$$
h_n(t) = c_0(t) u_1 u_2 \dots u_n.
$$
But condition \re{bh} is equivalent to \re{phi_H} and hence
$V_N(t)=b_N(t)$ and hence we have the 
\begin{pr}
Under initial condition \re{u0=t} the solution $b_N(t)$ of the corresponding Toda chain equations \re{Toda_eq} coincides with the unique rational solutions of the Painlev\'e-II equation with $\alpha=N+1/2$.
\end{pr}
 Note that the Hankel determinants $H_n(t)$ in this case coincide with so-called Yablonskii-Vorob'ev
polynomials \cite{JKM}. These polynomials were introduced in order to describe all rational solutions of
Painlev\'e-II equation. In our approach these polynomials appear quite naturally
under the simplest choice of the linear potential in the Schr\"odinger equation.

Corresponding Schr\"odinger equation \re{shcr} 
\be
\ddot \psi + (t -z^2/4) \psi =0 \lab{SE_VY} \ee
describes a quantum particle in the linear potential (say, in the uniform gravity field near the Earth surface) \cite{LL}.
Its general solution is well known
\be
\psi(z;t) = Q_1(z) Ai (z^2/4-t) +  Q_2(z) Bi(z^2/4-t), \lab{sol_Ai} \ee
where $Ai(x)$ and $Bi(x)$ are the standard Airy functions \cite{AbSt} and $Q_1(z), Q_2(z)$ arbitrary functions in $z$.
From asymptotic behavior at $z \to \infty$ \cite{AbSt} we can conclude that in \re{sol_Ai} necessarily $Q_1(z) \equiv 0$.
The term $Q_2(z)$ can be arbitrary and we can put $Q_2(z) =1$ without loss of generality.

For the Stieltjes function we then have from \re{F_psi_S}
\be
F(z;t) = \frac{\dot \psi(z;t)}{\psi(z;t)} + z/2 = -\frac{Bi'(\zeta)}{Bi(\zeta)} + z/2, \quad \zeta = z^2/4-t. \lab{F_Bi} \ee
Thus the Stieltjes function $F(z;t)$ has a simple explicit expression in terms of logarithmic derivative of the Airy function $Bi(x)$.
In a slightly different form this result was obtained in \cite{IKN}. In our approach this result follows naturally from the
Toda-Schr\"odinger correspondence.

On the other hand, it can be shown that these
orthogonal polynomials belong to a special type of the
Laguerre-Hahn polynomials.

Indeed, the Laguerre-Hahn orthogonal polynomials are defined through their Stieltjes function $F(z)$
satisfying the Riccati equation \cite{MagR}
\be
A(z) F'(z) = B(z) F^2(z) + C(z) F(z) + D(z), \lab{LHF} \ee
where $A(z), B(z), C(z), D(z)$ are polynomials in $z$ having no common zeros.

From results \cite{IKN} one can obtain that the Stieltjes function corresponding to the moments with initial condition $c_0=u_0=t$
satisfies the Riccati equation (in \cite{IKN} this Riccati equation appeared in a slightly different form due to initial choice of the generating function for the moments $c_n(t)$)
\be
 2F'(z;t) =  z F^2(z;t) - z^2 F(z;t) + tz+1. \lab{VY_LH} \ee
We thus see that our orthogonal polynomials indeed belong to the Laguerre-Hahn class with $A(z)=2, B(z)= z, C(z) =-z^2, D(z;t) = tz+1$.

%\vspace{15mm}

\section*{Acknowledgements}
%{\bf \Large Acknowledgements.} 
The authors are grateful to L.~Vinet for discussion. AZ thanks Kyoto University for hospitality.

%\vspace{15mm}

\bibliographystyle{amsplain}

\end{document}